\documentclass[12pt]{iopart}

\usepackage{amssymb}

\jl{1}

\def\rlx{\relax\leavevmode}
\def\IZ{\rlx\hbox{\small \sf Z\kern-.4em Z}}
\def\IR{\rlx\hbox{\rm I\kern-.18em R}}
\def\ID{\rlx\hbox{\rm I\kern-.18em D}}
\def\IC{\rlx\hbox{\,$\inbar\kern-.3em{\rm C}$}}
\def\IN{\rlx\hbox{\rm I\kern-.18em N}}
\def\one{\hbox{{1}\kern-.25em\hbox{l}}}

\def\half{\frac{1}{2}}
\def\ra{\rightarrow}
\def\beq{\begin{equation}}
\def\eeq{\end{equation}}
\def\bea{\begin{eqnarray}}
\def\eea{\end{eqnarray}}
\def\ber{\begin{array}}
\def\eer{\end{array}}
\def\xmu{x^{\mu}}
\def\xnu{x^{\nu}}
\def\eps{\varepsilon}
\def\ie{{\it i.e.\ }}

\def\half{\frac{1}{2}}

\newcommand{\smbox}[1]{{\mbox {\scriptsize #1}}}
\def\gbl{\left[\!\left[}
\def\gbr{\right]\!\right]}

\def\Ng{N_{\smbox{gh}}}

\newcommand{\eqn}[1]{(\ref{#1})}
\def\F{{\cal F}}
\def\L{{\cal L}}

\def\D21a{D(2,1;\alpha)}

\begin{document}

\title{Generalised scalar particle quantisation in $1+1$ dimensions and $D(2,1;\alpha)$}

\author{S P Corney\dag, P D Jarvis\dag, I Tsohantjis\dag and D S McAnally\ddag}
\address{\dag School of Mathematics and Physics, University of Tasmania
 GPO Box 252-21, Hobart Tas 7001, Australia }

\address{\ddag Department of Mathematics and Statistics, University of Melbourne, Parkville Victoria 3010, Australia}

\begin{abstract}
The exceptional superalgebra $\D21a$ has been classified as a candidate conformal supersymmetry algera in two dimensions. We propose an alternative interpretation of it as an extended BFV-BRST quantisation superalgebra in $2D$ ($D(2,1;1) \simeq osp(2,2|2)$).  A superfield realization is presented wherein the standard extended  phase space coordinates can be identified. The physical states are studied via the cohomology of the BRST operator. Finally we reverse engineer a classical action corresponding to the algebraic model we have constructed, and identify the Lagrangian equations of motion.
\end{abstract}

\noindent
UTAS-PHYS-00-15 \\
\submitted

\maketitle

\section{Introduction and main results}

In two previous papers we have examined the covariant BFV-BRST quantisation of the scalar \cite{scalar} and spinning \cite{spinning} particle respectively. In both these papers we started with a physical model for the system, whose quantisation was shown to obey the $iosp(d,2/2)$ algebra. In this paper we take an algebraic approach; as $osp(d,2/2)$ is a member of the class of classical simple Lie superalgebras, by an appropriate generalisation it should be possible to extend the  quantisation superalgebra $iosp(d,2/2)$ into a more general classical simple Lie superalgebra. The motivation behind this is the need for a characterisation of admissible spacetime `BFV-BRST extended' supersymmetries in various dimensions. Here we demonstrate this by studying the particular case of $d=2$, which leads to the quantisation of two-dimensional relativistic particles in the exceptional superalgebra $\D21a$.

In this section we briefly define and review the properties of the exceptional superalgebra $\D21a$. In section \ref{d21aquant} we shall construct superfield representations of the BFV-BRST quantisation superalgebra corresponding to $\D21a$ and study the physical states via the BRST operator. This will be done using only the algebraic structure as a guide (\ie no physical model). Finally, in section \ref{d21aclass} we shall reverse-engineer a classical action corresponding to the algebraic model we have constructed, and identify the corresponding Lagrangian equations of motion. A preliminary version of the results contained in this paper was published in \cite{d21agrp22}.

The classical simple Lie superalgebras consist of the $spl(m/n)$ and the $osp(m/2n)$ families, the strange series $P(n)$ and $Q(n)$ and the exceptional algebras $F(4)$, $G(3)$ and $\D21a$.  Comprehensive definitions and descriptions of these algebras can be found in several places, see for example \cite{kac, scheunert, farmerPhD,frappat}. A study of the $\D21a$ algebras, including a detailed analysis of their finite and infinite-dimensional irreducible representations has been carried out by Van der Jeugt \cite{jeugt}. The explicit supercommutation relations of the $\D21a$ superalgebras are given in \cite{scheunert2}.

The algebras $\D21a$ are a one-parameter family of 17-dimensional non-isomorphic Lie superalgebras. For the special case of $\alpha =1$ we have $D(2,1;1) \cong D(2,1) \cong osp(4,2)$. It is through this special case that we seek to generalise the BFV-BRST quantisation algebra. This aspect will be discussed in more detail in the next section.

The even part of the real superalgebra $\D21a$ is the 9-dimensional non-compact form $sl(2,\IR) + sl(2,\IR) + sl(2,\IR)$, whilst the odd part (of dimension 8) is the spinorial representation $(2,2,2)$ of the even part. The parameter $\alpha$ appears only in the anti-commutation relations among the components of the tensor products (\ie the odd components). In terms of the vectors $\eps_1, \eps_2, \eps_3$ such that $\eps_1^2 = -(1 + \alpha)/2, \; \eps_2^2 = 1/2, \; \eps_3^2 = \alpha/2
$ and $\eps_i \cdot \eps_j =0$ if $i\neq j$, the root system $\Delta = \Delta_{\bar{0}} \cup\Delta_{\bar{1}}$ is given by
\[
\begin{array}{ccc}
\Delta_{\bar{0}} = \left\{ \pm 2\eps_i \right\} & {\mbox and} & \Delta_{\bar{1}} = \left\{ \pm\eps_1 \pm\eps_2\pm\eps_3 \right\}.
\end{array}
\]

In \cite{gunaydin}, G\"{u}naydin studied $\D21a$  considered as the superconformal symmetry group of an family of exotic superspaces in two dimensions defined by the one-parameter family of Jordan superalgebras $JD(2/2)_\alpha$. In this paper he also derived the full super-differential operators representing the actions of $\D21a$ on the exotic superspaces. Here we similarly derive a superfield realisation, however it is over a different superspace.

\section{The quantised $\D21a$ particle}
\label{d21aquant}

\subsection{Preliminaries}

The BFV-BRST quantisation of relativistic systems provides a cohomological resolution of irreducible unitary representations (unirreps) of space-time symmetries. Moreover these unirreps appear to be associated with constructions of $iosp(d,2/2)$ for relativistic particles in flat spacetime, as can be seen in \cite{scalar} and \cite{spinning}. In this paper, however, we do not invoke translations as additional generators and so the algebra reduces to $osp(d,2/2)$. Here we follow an algebraic approach, and so require a classification of admissible `quantisation superalgebras' in various dimensions. Some examples of such algebras for Minkowski space $(d-1,1)$ \cite{gunaydin, gunaydin2} are $D(2,1;\alpha)$ in $d = 2$ (note that $\alpha =1$ corresponds to $osp(2,2/2)$); in $d =3$ we have $osp(3,2/2)$ which corresponds to anti de Sitter symmetry (which may thus be relevant to anyon quantisation), and for $d = 3+1$ we get conformal symmetry of 4 dimensional spacetime, and super unitary superalgebras, as possible alternative quantisation superalgebras.

In order to detail our construction for the d=2 case, we firstly outline the scalar particle quantisation in generic $d$. In second order form \cite{govbk} the action is
\beq
S = m \int_{\tau_{i}}^{\tau_{f}} d\!\tau \sqrt{\frac{d\xmu}{d\tau}\frac{d\xnu}{d\tau} \eta_{\mu\nu}},
\label{d2eq:act}
\eeq
which leads to the canonical momenta
\[
p_\mu = \frac{\partial \L}{\partial \dot{x}^\mu} = m \frac{\dot{x}_\mu}{\dot{x}^2}. 
\]
In accordance with the BRST prescription, we enlarge the phase space by treating the Lagrange multiplier $\lambda$ as a dynamical variable, along with an associated vanishing conjugate momentum $\pi_\lambda$. The system has two constraints; the first being the mass-shell condition 
\[
\phi_1 = p^\mu p_\mu - m^2 = 0,
\] 
and the second the aforementioned momenta conjugate to the Lagrange multiplier.
The Poisson brackets are the usual ones: $\{p_\mu,x^\nu\} = \delta^\nu_\mu, \; \{\lambda, \pi_\lambda \} = 1$.

The extended action \eqn{d2eq:act} is invariant under the following infinitesimal gauge transformations
\begin{eqnarray}
\delta \lambda &=& \dot{\eps}, \nonumber\\
\delta x^{\mu} &=& \{ x^{\mu},\eps \phi_1 \} = 2 \eps p^{\mu},
 \nonumber\\
\delta p^{\mu} &=& \{ p^{\mu},\eps \phi_1 \} = 0, \nonumber
\end{eqnarray}
with $\eps(\tau)$ being an arbitrary (dimensionless) infinitesimal function such that $\eps(\tau_i) = \eps(\tau_f) = 0$.

In order to derive the equations of motion for the action \eqn{d2eq:act} it is necessary to choose a particular gauge fixing condition. From the transformations of the einbein (which arises explicitly in the first-order formulation) under world-line diffeomorphisms, we have
\[
\frac{d \tilde{\tau}}{d \tau} = \frac{e(\tau)}{\tilde{e}(\tilde{\tau})}.
\]
It is thus always possible to find a parametrisation $\tilde{\tau}(\tau)$ such that $\tilde{e}(\tilde{\tau})$ is a constant. In the infinitesimal case we can write
\[
\tau = \tilde{\tau} + \frac{\eps(\tilde{\tau})}{e(\tau)},
\]
and so we have to this order
\[
\delta e = \tilde{e}(\tau) - e(\tau) = \dot{\eps}.
\]
Using this relationship allows us to identify the Lagrange multiplier $\lambda(\tau)$ with the einbein $e(\tau)$.

As in our two previous papers \cite{scalar, spinning} it is necessary to place two restrictions on the system so as to arrive at the particle quantisation corresponding to the superalgebraic prescription. Firstly, we take gauge fixing with respect to gauge transformations in the connected component of the group. This is equivalent to limiting the quantisation of $\lambda$ to the half line ($\IR^+$ or $\IR^-$). Secondly we take as a first class constraint $\phi_2 = \lambda \pi_\lambda$ (rather than the usual $\phi_2 = \pi_\lambda$ used in the standard construction). Note that $\phi_2 = \pi_\lambda$ is regular in the sense of Govaerts \cite{govbk} provided $\lambda \neq 0$.

If we now extend the phase space and in the same manner to that given in \cite{scalar} then we can write the standard BRST operator
\[
\Omega = \eta^1 \phi_1 + \eta^2 \phi_2.
\]
With a gauge fixing function defined as $\F = -\half \lambda \rho_2$  the Hamiltonian can be calculated as
\beq
H = \{ \F,\Omega \} = - \half \lambda \left( \eta^1 \rho_2 + p^\mu p_\mu - m^2\right).
\eeq
Similarly, quantising the system via the standard Schr\"{o}dinger representation we have the operators
\beq
X^\mu, P_\nu, \;\; \lambda, \pi_\lambda, \;\; Q_\alpha, X_\beta, \;\;P_+, X_-, \;P_- = H, X_+ = \tau.
\label{d2:opers}
\eeq
The non-zero commutation relations between these operators are
\beq
\begin{array}{ccc}
\left[X_\mu ,P_\nu \right]=-i\eta_{\mu\nu}, &
\left\{X_\alpha,Q_\beta \right\} = i\eps_{\alpha\beta},
&\left[X_-,P_+\right]=i,  \\
\left[X_-,P_-\right] = - iP_+^{-1} P_-, &  \left[X_\alpha
,P_-\right] =iP_+^{-1} Q_\alpha , &
\left[X_\mu,P_-\right]=iP_+^{-1}P_\mu.
\end{array}
\label{eq:rawmat1}
\eeq
Finally we find that the operators $J_{AB}$ defined as 
\beq
\begin{array}{cc}
J_{\mu-} = X_\mu P_- - X_- P_\mu, & L_{\mu\alpha} = X_\mu Q_\alpha - \theta_\alpha P_\mu, \\
J_{\mu\nu} = X_\mu P_\nu - X_\nu P_\mu, & J_{+\mu} = X_+ P_\mu - X_\mu P_+, \\
L_{+\alpha} = X_+ Q_\alpha - \theta_\alpha P_+, & K_{\alpha\beta} = \theta_\alpha Q_\beta + \theta_\beta Q_\alpha, \\
J_{+-} = X_- P_+ - X_+ P_-, & L_{\alpha-} = \theta_\alpha P_- - X_- Q_\alpha,
\end{array}
\label{eq:js}
\eeq
generate the inhomogeneous orthosymplectic superalgebra $iosp(d,2/2)$. The graded commutation relations of this algebra in tensor were given in \cite{pdjhsg}, and are repeated here as
\bea
\gbl J_{MN},J_{PQ} \gbr &=& i(\eta_{NQ}J_{MP} - [NP] \eta_{NP} J_{MQ} \nonumber \\
&&-[MN] [MP] \eta_{MP} J_{NQ} + [PQ][MN][MQ] \eta_{MQ} J_{NP}).
\label{eq:homopart}
\eea
\beq
\gbl J_{MN},P_{L} \gbr = i(\eta_{LN} P_M - [MN]\eta_{LM} P_N,
\label{eq:inhomopart}
\eeq
where the sign factors $[MN] \equiv (-1)^{m \cdot n}$ are both $-1$ when both indicies are Fermionic.

In the $d=2$ case, in order to extend $osp(2,2/2)$ to $D(2,1;\alpha)$, we must modify three of the anticommutation relations given in \eqn{eq:homopart} and \eqn{eq:inhomopart} (with the rest remaining the same). The new relations are
\bea
\{L_{\mu\alpha},L_{\nu\beta} \} &=& \eps_{\alpha\beta}\eps_{\mu\nu}(J + A J_{+-}) - \eta_{\mu\nu}K_{\alpha\beta} \nonumber\\
\{ L_{\mu\alpha},L_{\beta\pm} \} &=& -\eps_{\alpha\beta}(J_{\mu\pm} \pm B_\pm \eps_\mu ^\nu J_{\nu\pm}), \label{d2eq:firstjs}\\
\{ L_{\alpha\pm},L_{\beta\mp} \} &=& \pm \eps_{\alpha\beta} (J_{+-} \pm C_\pm J) - K_{\alpha\beta}. \nonumber
\eea
For simplicity, we recognise that $J_{\mu\nu}$ is anti-symmetric. This allows us to define the operator $J$ by
\beq
J_{\mu\nu} = \eps_{\mu\nu} J.
\eeq
Taking the super-Jacobi identity on $L_{\mu\alpha}, L_{\nu\beta}$ and $L_{\gamma\pm}$ it is straightforward to show that
\[
B_{\pm} = \frac{\mp A}{{\rm det}(\eta)}.
\]
Taking the super-Jacobi identity on $L_{\mu\alpha}, L_{\mu\pm}$ and $L_{\gamma\mp}$ we can show
\[
C_{\pm} = \frac{-A}{{\rm det}(\eta)}.
\]
Through the use of Cartan generators and weight operators we can eliminate all but one of $A,B_\pm,C_\pm$ and relate them back to the $\alpha$ parameter in $D(2,1;\alpha)$. This results in new generators
\bea
\tilde{J} &=& J + a J_{+-}, \\
\tilde{J}_{+-} &=& J_{+-} + a J, \nonumber
\eea
with the single parameter $a$, given by
\beq
a = \frac{1 - \alpha}{1+ \alpha}.
\eeq
The modified anti-commutation relations \eqn{d2eq:firstjs} can now be written
\bea
\{L_{\mu\alpha},L_{\nu\beta} \} &=& \eps_{\alpha\beta}\eps_{\mu\nu}\tilde{J} - \eta_{\mu\nu}K_{\alpha\beta} \nonumber\\
\{ L_{\mu\alpha},L_{\beta\pm} \} &=& -\eps_{\alpha\beta}(J_{\mu\pm} \pm a \eps_\mu ^\nu J_{\nu\pm}), \label{d2eq:secondjs}\\
\{ L_{\alpha\pm},L_{\beta\mp} \} &=& \pm \eps_{\alpha\beta} \tilde{J}_{+-} - K_{\alpha\beta}. \nonumber
\eea
Note that the odd generators close compactly on $\tilde{J}, \tilde{J}_{+-}$ (although at the expense of more complicated commutation brackets). The invariant bilinear form on $D(2,1;\alpha)$ is
\begin{eqnarray*}
(J,J) &=& 1, \\
(J,J_{+-}) &=& (J_{+-},J) = -a, \\
(J_{+-},J_{+-}) &=& 1, \\
(J_{\mu\pm},J_{\nu\pm}) &=& -\eta_{\mu\nu} \pm a\eps_{\mu\nu}, \\
(K_{\alpha\beta},K_{\gamma\delta}) &=& (1-a^2) \left( \eps_{\alpha\gamma} \eps_{\beta\delta} + \eps_{\alpha\delta} \eps_{\beta\gamma} \right), \\
(L_{\mu\alpha}, L_{\nu\beta}) &=& (1-a^2) \eta_{\mu\nu}\eps_{\alpha\beta}, \\
(L_{\alpha\pm}, L_{\beta\mp}) &=& (1-a^2) \eps_{\alpha\beta}.
\end{eqnarray*}
And the Casimir $C$ can be written 
\beq
C = C_1 + C_2,
\eeq
where
\bea
C_1 &=& -J^2 - J_{+-}^2 + \left\{J^\mu_+,J_{\mu-}\right\} - \half K^{\alpha\beta} K_{\alpha\beta} - L^{\mu\alpha} L_{\mu\alpha} - \left[ L^\alpha_+, L_{\alpha-} \right], \nonumber \\
C_2 &=& -a\left\{J,J_{+-} \right\} - a\eps^{\mu\nu} \left\{J_{\mu+}, J_{\nu-} \right\}. \nonumber
\eea

\subsection{Superfield Realisation}

For the $\D21a$ particle we do not have a physical model such as that laid out between equations \eqn{d2eq:act} to \eqn{d2:opers}, and so must use the algebraic structure as our only guide. We regard $\D21a$ as a generalisation of $osp(d,2/2)$ (for the $d=2$ case) and seek to find a superfield realisation which is equivalent to the case for the scalar relativistic particle for $\alpha = 1$ and $d=2$.

In the generic $d$-dimensional $osp(d,2/2)$ case we can define the homogeneous manifold
\[
{\mathcal M} = OSp(d,2/2) / G_0,
\]
where $G_0$ is the stability group
\[
G_0 = OSp(d-1,2/2) \wedge {\mathcal N},
\]
and
\bea
OSp(d-1,2/2) &=& \langle J_{\mu\nu}, L_{\mu\alpha}, K_{\alpha\beta} \rangle, \nonumber \\
{\mathcal N} &=& \langle J_{\mu -}, L_{\alpha -} \rangle.
\label{d2eq:OspN}
\eea
For one parameter subgroups $g(t)$ with generator $A$,  the standard superfield realisation leads to generators $\hat{A}$ acting on functions $\phi$ over ${\mathcal M}$, defined by
\beq
\hat{A}\phi(x) = \left(\frac{d}{dt}\phi(g(t)^{-1}x) \right)_{t=0},
\label{d2eq:A}
\eeq
where $x \in {\mathcal M}$, 
\[
 x = (q^\mu ,\eta^\alpha , \phi) \leftrightarrow \exp (q^\mu J_{\mu+} + \eta^\alpha L_{\alpha+}) \exp(\phi J_{+-})G_0, 
\]
represents the coset.

In a similar fashion, for $\D21a$ we define the homogeneous manifold and stability group as
\bea
{\mathcal M} &=& D(2,1;\alpha) / \tilde{G}_0, \nonumber \\
\tilde{G}_0 &=& Osp(1,1/2) \wedge {\mathcal N}, \nonumber
\eea
where now $OSp(1,1/2) = \langle \tilde{J}, L_{\mu\alpha}, K_{\alpha\beta} \rangle$ and ${\mathcal N}$ is unchanged from \eqn{d2eq:OspN}. This leads to generators $\hat{A}$ defined as in \eqn{d2eq:A}, except now the coset representatives $x \in {\mathcal M}$ are given by
\[
x \equiv (q^\mu, \eta^\alpha, \phi) \leftrightarrow \exp(q^\mu J_{\mu+} + \eta^\alpha L_{\alpha+}) \exp(\phi\tilde{J}_{+-}) \tilde{G}_0.
\]

The superfield realisation for $\D21a$ is computed in the standard way following \eqn{d2eq:A}. For example, it is clear that the even generators $J_{\mu\nu}$ associated with group elements $g(\epsilon) = \exp(\epsilon^\mu J_{\mu\nu})$, simply induce translations in the co-ordinates $q^\mu, \; \hat{J}_{\mu +} = - \partial / \partial q^\mu$. For later comparison, we re-scale the variables as follows:
\[
p^\mu = \lambda^{-1} q^\mu, \;\; \theta^\alpha = \lambda^{-1} \eta^\alpha, \;\;\lambda = e^\phi, (\lambda >0).
\]
Then, finally we have
\[
J_{\mu+} = -\lambda^{-1} \frac{\partial}{\partial p^\mu}.
\]
In \ref{appenA} we explicitly evaluate $J_{\mu+}$, as well as a further two generators, with the understanding that the remainder are done in a similar fashion.

The full set of generators for the superfield realisation of $\D21a$ is 
\bea
J_{\mu +} &=& - \lambda^{-1} \frac{\partial}{\partial p^{\mu}}, \nonumber\\
L_{\alpha +} &=& - \lambda^{-1} \frac{\partial}{\partial \theta^{\alpha}},
\nonumber\\
L_{\alpha -} &=& \frac{1}{2} \lambda (p^{\nu} p_{\nu}
+ \theta^{\beta} \theta_{\beta}) \frac{\partial}{\partial \theta^{\alpha}}
- \theta_{\alpha} \lambda^2 \frac{\partial}{\partial \lambda}
- a \lambda \theta_{\alpha} p^{\mu} \varepsilon_{\mu}{}^{\nu}
\frac{\partial}{\partial p^{\nu}}, \nonumber\\
L_{\mu \alpha} &=& p_{\mu} \frac{\partial}{\partial \theta^{\alpha}}
- \theta_{\alpha} \frac{\partial}{\partial p^{\mu}} - a \theta_{\alpha}
\varepsilon_{\mu}{}^{\nu} \frac{\partial}{\partial p^{\nu}}, \nonumber\\
K_{\alpha \beta} &=& \theta_{\alpha} \frac{\partial}{\partial \theta^{\beta}}
+ \theta_{\beta} \frac{\partial}{\partial \theta^{\alpha}}, \label{eq:fulljs} \\
J &=& - p^{\mu} \varepsilon_{\mu}{}^{\nu} \frac{\partial}{\partial p^{\nu}}
+ \frac{a}{1-a^2} \left(\lambda \frac{\partial}{\partial \lambda}
- p^{\mu} \frac{\partial}{\partial p^{\mu}} - \theta^{\alpha}
\frac{\partial}{\partial \theta^{\alpha}} \right), \nonumber\\
J_{+-} &=& - \lambda \frac{\partial}{\partial \lambda}
- \frac{a^2}{1-a^2} \left(\lambda \frac{\partial}{\partial \lambda}
- p^{\mu} \frac{\partial}{\partial p^{\mu}} - \theta^{\alpha}
\frac{\partial}{\partial \theta^{\alpha}} \right), \nonumber \\
J_{\mu -} &=& \frac{1}{2} \lambda\theta^{\alpha} \theta_{\alpha} 
\frac{\partial}{\partial p^{\mu}} 
+ \frac{1}{2} \lambda a \varepsilon_{\mu}{}^{\nu} \theta^{\alpha} 
\theta_{\alpha} \frac{\partial}{\partial p^{\nu}} 
+ \frac{1}{2} \lambda \varepsilon_{\mu \nu} \varepsilon_{\rho}{}^{\sigma} 
p^{\nu} p^{\rho} \frac{\partial}{\partial p^{\sigma}}  \nonumber\\
&&\mbox{\quad}- \lambda^2 p_{\mu} \frac{\partial}{\partial \lambda}
+ \frac{1}{2} \lambda p_{\mu} p^{\nu} \frac{\partial}{\partial p^{\nu}} 
\nonumber\\
&& \mbox{\quad}- \frac{\lambda a (a p_{\mu} + \varepsilon_{\mu \rho} p^{\rho})}
{1-a^2}
\left(\lambda \frac{\partial}{\partial \lambda} 
- p^{\nu} \frac{\partial}{\partial p^{\nu}} 
- \theta^{\alpha} \frac{\partial}{\partial \theta^{\alpha}} \right). \nonumber 
\eea

If we compare this realisation with that obtained for $osp(d,2/2)$ (see \cite{scalar}) the similarities are evident (although the realisation of $J_{\mu-}$ requires some attention). In fact, if we allow $\alpha \ra 1$ (and thus $a \ra 0$), which corresponds to $\D21a \cong osp(2,2/2)$, then it can easily be seen that the above relations are in fact identical to those obtained using the standard superfield \cite{scalar} for the massless case. In \ref{appenB} the closure of these generators on the $\D21a$ algebra is illustrated for the case of the anti-commutator $\{L_{\mu\alpha}, L_{\nu\beta} \} = \eps_{\alpha\beta} \eps_{\mu\nu} \tilde{J} - \eta_{\mu\nu} K_{\alpha\beta}$.

\subsection{Physical States}

The BRST operator for the $\D21a$ model can be constructed by considering two linearly independent spinors $\eta^\alpha$ and $\eta'^\alpha$ which obey the condition $\eta^\alpha \eta'_\alpha = 1$, for example
\[
\eta^\alpha = \frac{1}{\sqrt{2}} \left( \begin{array}{c} 1 \\ 1 \end{array} \right), \;\; \eta'^\alpha = \frac{1}{\sqrt{2}} \left( \begin{array}{c} -1 \\ 1 \end{array} \right).
\]
We define the ghost number operator in the usual way, 
\beq
\Ng = \eta^\alpha \eta'^\beta K_{\alpha\beta}= (\eta \cdot \theta)(\eta'\cdot \partial) + (\eta' \cdot \partial)(\eta \cdot \partial),
\label{d2eq:ghno}
\eeq
where $\partial_\alpha = \partial / \partial \theta^\alpha$. Similarly we can define the BRST operator as
\bea
\Omega &=& \eta^\alpha L_{\alpha-}, \nonumber \\
&=& \eta^\alpha \left(\frac{1}{2} \lambda (p^{\nu} p_{\nu} + \theta^{\beta} \theta_{\beta}) \frac{\partial}{\partial \theta^{\alpha}} - \theta_{\alpha} \lambda^2 \frac{\partial}{\partial \lambda} - a \lambda \theta_{\alpha} p^{\mu} \varepsilon_{\mu}{}^{\nu} \frac{\partial}{\partial p^{\nu}} \right).
\eea
The physical states can be calculated by considering the effect of $\Omega$ on a superfield
\beq
\psi = A + \theta^\alpha \chi_\alpha + \half \theta^\alpha \theta_\alpha B
\label{d2eq:psi}
\eeq
(for more detail see \cite{thesis}). Explicitly we can write
\bea
\Omega \psi &=& \frac{1}{2p_+} p \cdot p (\eta \cdot \chi) 
+ \eta^{\alpha} \theta_{\alpha} \left( \frac{1}{2 p_+} B 
+ \left[ \frac{\partial}{\partial p_+} 
- \frac{a}{p_+} (p^\mu \eps_{\mu\nu} \partial_{p_\mu}) \right] A \right) \nonumber\\
&& + \frac{1}{2} \theta^{\alpha} \theta_{\alpha} 
\left(-\frac{\partial}{\partial p_+} + \frac{1}{p_+}(1+ap^\mu \eps_{\mu\nu}
\partial_{p_\nu}) \right)(\eta \cdot \chi), \label{d2eq:omegapsi}
\eea
and imposing the conditions
\[
\Omega \psi = 0, \;\; \psi \neq \Omega \psi' \mbox{ and } \Ng \psi = \ell \psi,
\]
for the maximal eigenvalue $\ell = 1$ (in accordance with \cite{thesis}).

Comparing equations \eqn{d2eq:psi} and \eqn{d2eq:omegapsi} we see that the components $A, \chi_\alpha$ and $B$ of $\psi$ are defined up to addition of functions corresponding to coefficients in \eqn{d2eq:omegapsi}. Imposing the first condition above, we see that the $\eta^\alpha\theta_\alpha$ coefficient determines simply the $p_+$-dependence of $A$ in terms of some unknown $B$ (which is itself determined up to a $p_+$-derivative of some function).

Looking at $\eta^\alpha \chi_\alpha$, we find, respectively, from ${\mathcal O}(\theta^0)$ and ${\mathcal O}(\theta^2)$
\bea
\frac{1}{2 p_+} (p\cdot p) (\eta^\alpha \chi_\alpha) &=& 0, \label{d2eq:3cross} \\
\left( \frac{\partial}{\partial p_+} - \frac{1 + a p^\mu \eps_{\mu\nu} \partial^\nu}{p_+} \right) (\eta^\alpha \chi_\alpha) &=& 0. \label{d2eq:3star}
\eea
We can write $p^\mu$ in component form as
\bea
p^\mu_R &=& \frac{1}{\sqrt{2}} (p^\mu + {\eps^\mu}_\nu p^\nu), \nonumber \\
p^\mu_L &=& \frac{1}{\sqrt{2}} (p^\mu - {\eps^\mu}_\nu p^\nu). \nonumber
\eea
Assuming that $\eta^\alpha \chi_\alpha = \phi(p)$, where $\phi(p)$ is a function of the form
\[
\phi(p) = p_+ \varphi\left( (1 + a \ln p_+) p_R, (1 - a \ln p_+) p_L \right),
\]
we have
\[
\frac{\partial}{\partial p_+} \phi(p) = \frac{1}{p_+} \left( 1 + a( p^\mu {\eps_\mu}^\nu \frac{\partial}{\partial p^\nu} ) \right) \phi,
\]
as in \eqn{d2eq:3star}. Hence the given form of $\phi(p)$ solves equation \eqn{d2eq:3star}.

Enforcing \eqn{d2eq:3cross} gives that $\phi(p)$ satisfies $p\cdot p = 0$, or 
\[
\frac{1}{2 p_+} p_R \cdot p_L p_+ \varphi(\zeta^+ p_R, \zeta^- p_L) = 0,
\]
where $\zeta^\pm = 1 \pm a \ln p_+$. Equivalently, in Fourier space the constraints are solved by the physical states
\[
\varphi(x_R,x_L) \equiv \int (\chi \cdot \psi) e^{-i p \cdot x} dx,
\]
that satisfy
\beq
\frac{\partial}{\partial x_R} \frac{\partial}{\partial x_L} \varphi\left( \frac{x_R}{\zeta^+}, \frac{x_L}{\zeta^-} \right) = 0.
\eeq 
Moreover, if we assume $H \psi = 0$ ($H$ the Hamiltonian) on the physical states and we assume the Schr\"{o}dinger equation
\[
H = -i\frac{d}{d \tau},
\]
then these functions are independent of $\tau$.

Finally, by once again employing the triviality of $Im \Omega$ for the maximal ghost number we can show that the physical state is unique, as there is no function $\psi'$ such that $\psi = \Omega \psi'$ lies in the same cohomology as $\psi$.

The system we have constructed can be interpreted as the `quantisation'  of a classical `$\D21a$' particle. The $p_+$-dependence on physical wavefunctions provides indirect evidence that the model involves a more subtle implementation of world line diffeomorphisms than usual. Note that the two-dimensional case has the unique property that Lorentz invariance is not broken. The metric $x^\mu x^\nu \eta_{\mu\nu} = x_R x_L$ is still a world-line scalar if $x_R,x_L$ transform as densities:
\[
x'_{R,L} (\tau') d \tau'^{\pm a} = x_{R,L}(\tau) d \tau ^{\pm a}.
\]
Corresponding covariant actions may be responsible (after gauge fixing) for the $p_+$-scaling behaviour\footnote{The gauge equivalence class of $\lambda$, or $e$, namely $\int_{\tau_{i}}^{\tau_{f}} e(\tau) d\,\tau$, is proportional to $\lambda$ in the present case $\dot{\lambda}=0$}.

\section{Classical Hamiltonian and action}
\label{d21aclass}

In the previous section we did not explicitly calculate the corresponding Hamiltonian $H = -\{ \F, \Omega \}$, nor did we specify a gauge fixing function $\F$. The reason behind this is simple, as we have no classical model with which to compare our quantised particle we do not have any guide as to what our quantised Hamiltonian should look like, and thus no guide as to which gauge fixing function $\F$ we should choose. In this section we postulate an $\F$ which leads to an acceptable looking Hamiltonian, and from there derive a classical action $S$. This is the action which defines the classical system which corresponds to the quantum system derived from the algebraic structure in section \ref{d21aquant}.

By definition the gauge fixing fuction $\F$ is Grassmann odd and has ghost number $-1$, thus it obeys the equation 
\beq
\gbl\Ng,\F \gbr = -\F.
\label{d2eq:ngf}
\eeq
As well as these constraints on $\F$, in the $\D21a$ system we must make sure that Hamiltonian generated is general enough to encompass the extended behaviour of the system (as compared with the corresponding $osp(2,2/2)$ system) and that in the limit $\alpha \ra 1$ it reduces to Hamiltonian for the $osp(2,2/2)$ system of \cite{scalar}.

Firstly we express the ghost number operator (see \eqn{d2eq:ghno}) as
\[
\Ng = \half (K_{22} - K_{11}) = \theta_2 \partial_2 - \theta_1 \partial_1.
\]
As a first guess at the gauge fixing function we choose $\F = \eta'^\alpha \theta_\alpha$. This function has ghost number $-1$ and is Grassmann odd, checking that it satisfies \eqn{d2eq:ngf} we get
\[
\gbl \Ng, \F \gbr = \frac{1}{\sqrt{2}} \{ \theta_2 \partial_2 - \theta_1 \partial_1, \theta_2 - \theta_1 \} = - \frac{1}{\sqrt{2}} ( \theta_2 - \theta_1 ) = - \F,
\]
as we desire. However this $\F$ falls down when we generate the corresponding Hamiltonian, which is found to be independent of $\alpha$. Thus the Hamiltonian generated by this gauge fixing function cannot reproduce the $\alpha$ dependent quantised system of section \ref{d21aquant}.

Our second choice for the gauge fixing function is $\F = \eta'^\alpha \partial_\alpha$. This function also satisfies the necessary conditions however it once again falls down when we generate the Hamiltonian, as this $H$ does not revert to the $osp(2,2/2)$ Hamiltonian as $\alpha \ra 1$. Thus we are led to choosing our gauge fixing function as a scalar combination of the two given above, \ie
\beq
\F = \frac{1}{\sqrt{2}} \left( (\theta_2 - \theta_1) + b (\partial_1 - \partial_2) \right),
\label{d2eq:f}
\eeq
where $b$ is an arbitrary scalar, and we have changed the overall sign of the second term. This $\F$ is Grassmann odd, has ghost number $-1$ and obeys equation \eqn{d2eq:ngf}. The corresponding Hamiltonian can now be calculated
\[
H =  \gbl \Omega, \F \gbr = \gbl \Omega, \F_\theta -b \F_\partial \gbr.
\]
\bea
\gbl \Omega, \F_\theta \gbr &=& \gbl \half \lambda p^2 ( \partial_1 + \partial_2) + \half \lambda \theta^\beta \theta_\beta (\partial_1 + \partial_2), \theta_2 - \theta_1 \gbr, \nonumber \\
&=& -\half \lambda ( p^2 + \theta^\beta \theta_\beta ), \nonumber
\eea
and
\bea
\gbl \Omega, -\F_\partial \gbr &=& \half \gbl - (\theta_1 + \theta_2) \lambda^2 \partial_\lambda - a \lambda (\theta_1 + \theta_2) p^\mu \eps_\mu^\nu \partial_\nu + \lambda \theta^\beta \theta_\beta (\partial_1 + \partial_2), \partial_a - \partial_2 \gbr, \nonumber \\
&=& \lambda^2 \partial_\lambda + a \lambda p^\mu \eps_\mu^\nu \partial_\nu + \half \lambda (\theta_1 - \theta_2)(\partial_1 + \partial_2). \nonumber
\eea
Thus the Hamiltonian for the $\D21a$ system is
\beq
H= -\half \lambda \left( p^2 + \theta^\beta \theta_\beta \right) + b \lambda^2 \partial_\lambda + a b \lambda p^\mu \eps_\mu^\nu \partial_\nu + \half b \lambda (\theta_1 - \theta_2)(\partial_1 + \partial_2),
\label{d2eq:hamil}
\eeq
where $\lambda = 1/p_+$. Notably this action is general enough so as to encompass the special behaviour of the $\D21a$ system (as $a = a(\alpha)$) and can reduce to the Hamiltonian for the massless scalar particle in the $osp(2,2/2)$ case of \cite{scalar}.

Having derived the Hamiltonian of the $\D21a$ system we now seek to calculate the corresponding classical action and Lagrangian $\L$. We do this by means of a Legendre transformation and the Hamiltonian equations of motion. Given the Hamiltonian, we can write the Lagrangian as
\beq
\L = \sum \dot{q} p - H(q,p),
\label{d2eq:lag}
\eeq
where $q,p$ are the generalised co-ordinates of $H$. $\dot{q}$ is calculated by means of the Hamiltonian equations of motion;
\bea
\dot{x}^\mu = \frac{\partial H}{\partial p_\mu} &=& \lambda (-p^\mu + ab \eps^\mu_\nu x^\nu), \nonumber \\
\dot{\lambda} = \frac{\partial H}{\partial \partial_\lambda} &=& b \lambda^2, \\
\dot{\theta^\alpha} = \frac{\partial H}{\partial \partial_\alpha} &=& \frac{1}{\sqrt{2}} \lambda b (\theta_1 - \theta_2) \eta^\alpha \partial_\alpha, \nonumber \\
\therefore\;\;\; \dot{\theta}^\alpha \partial_\alpha &=& \half \lambda b (\theta_1 - \theta_2)(\partial_1 + \partial_2). \nonumber
\eea
Note that here we have used 
\[
\eta^\alpha \partial_\alpha = \frac{1}{\sqrt{2}} ( \partial_1 + \partial_2).
\]
From the first of these equations we can write
\[
p_\mu = -\frac{\dot{x}_\mu}{\lambda} - ab \eps_{\mu\nu} x^\nu,
\]
and so, by substituting the above expressions into \eqn{d2eq:lag}, we get
\bea
\L &=& \dot{x}^\mu \left( -\frac{\dot{x}_\mu}{\lambda} - ab \eps_{\mu\nu} x^\nu \right) + b \lambda^2 \partial_\lambda + \half \lambda b (\theta_1 - \theta_2)(\partial_1 + \partial_2)  \nonumber\\
&& + \half \lambda \left( (\frac{-\dot{x}_\mu}{\lambda} - ab \eps_{\mu\nu} x^\nu)(\frac{-\dot{x}^\mu}{\lambda} - ab \eps^{\mu\nu} x_\nu) + \theta^\beta \theta_\beta \right) - b \lambda^2 \partial_\lambda\\
&& - a b \lambda  \left(\frac{\dot{x}^\mu}{\lambda} - ab \eps^\mu_\nu x^\nu\right) \eps_\mu^\nu x_\nu - \half b \lambda (\theta_1 - \theta_2)(\partial_1 + \partial_2). \nonumber 
\eea
Thus the Lagrangian can be written
\beq
\L = -\half \frac{\dot{x}^2}{\lambda} - \lambda \half(ab)^2 x^2 - ab \eps_{\mu\nu} \dot{x}^\mu x^\nu + \half \lambda \theta^\beta \theta_\beta.
\label{d2eq:fulllag}
\eeq

The classical Lagrangian (and action) corresponding to the quantum Hamiltonian \eqn{d2eq:hamil} should be free of ghosts (as they only arise in the extended phase space of the BFV-BRST construction). Likewise the canonical momentum conjugate to the Lagrange multiplier $\lambda$ should not be present. Thus we arrive at the classical action of the $\D21a$ system by decoupling the ghost sector from the action above, \ie only considering the bosonic part
\beq
S = \int_{\tau_i}^{\tau_f} d\tau \left[ -\half \frac{\dot{x}^2}{\lambda} - \lambda \half(ab)^2 x^2 - ab \eps_{\mu\nu} \dot{x}^\mu x^\nu \right].
\label{d2eq:action}
\eeq
By comparing this with the action given in \cite{scalar} we can see that \eqn{d2eq:action} corresponds to a massless scalar particle in a potential well. In fact if we ignore the last term in \eqn{d2eq:action} then we have arrived at the classical action of an oscillating massless particle (\ie where the potential is proportional to $x^2$). For further details of the oscillator in the classical or quantum case see \cite{goldstein, aitchison}. The final term of \eqn{d2eq:action} introduces a cross term between velocity and position. Comparing this term with the potential term in \cite{scalar}, we see that the cross term is similar to that produced by a homogeneous electromagnetic field $F_{\mu\nu} = ab \eps_{\mu\nu}$.

The action \eqn{d2eq:action} also satisfies the condition that as $a \ra 0$ (which is equivalent to $\alpha \ra 1$), $\L$ becomes the Lagrangian of the massless scalar particle. The constant $b$ is also important as it distinguishes between the parts of the action that arise from each of the two gauge fixing functions we tried earlier; $\F_\theta$ and $\F_\partial$. We can now set $b=1$ without affecting the behaviour of the particle.

For the sake of completeness, we shall identify the total covariant energy and angular momentum of the classical $\D21a$ particle, as well as calculating the Euler-Lagrange equations of motion. The total covariant energy is given by
\bea
P_\mu &=& \frac{\partial \L}{\partial \dot{x}^\mu}, \nonumber \\
&=& - \frac{\dot{x}_\mu}{\lambda} - ab \eps_{\mu\nu} x^\nu. \nonumber
\eea
The total angular momentum is
\bea
M_{\mu\nu} &=& P_\mu x_\nu - P_\nu x_\mu, \nonumber \\
&=& \frac{1}{\lambda} (\dot{x}_\nu x_\mu - \dot{x}_\mu x_\nu) + ab ( \eps_{\nu\rho} x_\mu - \eps_{\mu\rho} x_\nu ) x^\rho.
\eea
The Euler Lagrange equations of motion are
\bea
-\frac{\ddot{x}_\mu}{\lambda} + (ab)^2 \lambda x_\mu - 2 ab \eps_{\mu\nu} \dot{x}^\nu &=& 0, \\
\frac{\dot{x}^2}{2\lambda^2} - (ab)^2 x^2 &=& 0. \label{d2eq:constraint}
\eea
By virtue of the fact that the Lagrangian is independent of $\dot{\lambda}$, the second of these two equations is actually a constraint on the system. Hence, the Hessian of the Lagrange function vanishes identically, except for its components $\partial^2 \L / (\partial \dot{x}^\mu \partial \dot{x}^\nu)$ \cite{govbk}, thereby illustrating the singular nature of the action \eqn{d2eq:action}. Thus we can identify \eqn{d2eq:constraint} as a first class constraint of the $\D21a$ system.

Now that we have determined the classical action corresponding to the $\D21a$ particle it is possible to start the loop again, so to speak. That is, identify the first class constraints, extend the phase space and follow the BFV-BRST quantisation procedure in order to arrive at the quantised $\D21a$ particle. However we shall not do this; given that we correctly choose the gauge fixing condition for the scalar particle condition we would end up with exactly the same quantised system as system as that obtained in section \ref{d21aquant}. Secondly, the aim in this paper is to study the algebras of quantisation, whch we have already done for the $\D21a$ particle in the previous section.

\section{Conclusion}

In this paper we have shown that it is possible to extend the BFV-BRST quantisation algebra $iosp(d,2/2)$ in two dimensions into the more general classical simple Lie superalgebra $\D21a$. To do this we started without a classical physical model of a particle, and so relied entirely on the algebraic structure as our guide. In section \ref{d21aquant} we showed that the algebraic model that we had constructed was an admissible quantisation superalgebra, and so provided a quantisation of the corresponding classical system. In section \ref{d21aclass} we then calculated the classical action corresponding to the quantum system. If this action was used as a starting point, then the BFV-BRST process could be followed and the quantum system of equation \eqn{d21aquant} would be derived. 

An alternative (and equally valid) method of presenting this paper would have been to start with the classical action \eqn{d2eq:action} and from there proceed to quantise the system and demonstrate that it obeys a $\D21a$ quantisation superalgebra.

\ack
IT and DSM acknowledge the Australian Research Council for the award of a Fellowship, and Tony Bracken and the Centre for Mathematical Physics, University of Queensland, for support. Finally SC acknowledges support from an Australian Postgraduate Award. Correspondence with Jan Govaerts and Ian McArthurs is also gratefully acknowledged.

\appendix

\section{Calculations for superfield realisation}
\label{appenA}

In this appendix we explicitly calculate the superfield realisation for $\D21a$ by introducing formal group elements $g = e^{\phi \cdot F}$ for the generators $F_1, F_2, \ldots F_N$ and graded parameters $\phi_1, \phi_2, \ldots \phi_N$ and evaluate the product
\[
h \cdot g = e^{\epsilon \cdot F} e^{ \phi \cdot F},
\]
in order to find the product map $\mu(\epsilon, \phi)$ to first-order in $\epsilon$.

Thus we can calculate $J_{\mu+}$ as follows:

\noindent
Let $\epsilon = e^{\epsilon^\mu J_{\mu+}}$, then
\bea
\epsilon ^{-1} \cdot x &=&e^{-\epsilon^\mu J_{\mu+}} e^{q^\mu J_{\mu+} + \eta^\alpha L_{\alpha+}} e^{\phi\tilde{J}_{+-}}G_0, \nonumber \\
&=& e^{(q^\mu - \epsilon^\mu) J_{\mu+} + \eta^\alpha L_{\alpha+}} e^{\phi\tilde{J}_{+-}} G_0. \nonumber
\eea
So if we consider functions $f(q^\mu, \eta^\alpha, \phi)$, then
\beq
\epsilon f(q^\mu, \eta^\alpha, \phi) = f(q - \epsilon, \eta, \phi),
\eeq
and so
\[
\delta f = \epsilon f - f = - \epsilon^\mu \frac{\partial}{\partial q^\mu}f(q^\mu, \eta^\alpha, \phi) + \ldots.
\]
Hence from \eqn{d2eq:A} we can write
\beq
J_{\mu+} = - \frac{\partial}{\partial q^\mu}.
\eeq

We shall explicitly calculate a further two generators, with the understanding that the remainder can be derived in a similar fashion.

\noindent
Let $\xi = e^{\xi^\alpha L_{\alpha+}}$, thus
\bea
\xi^{-1} x &=& e^{-\xi^\alpha L_{\alpha -}}  e^{q^\mu J_{\mu+} + \eta^\alpha L_{\alpha+}} e^{\phi\tilde{J}_{+-}}G_0, \nonumber \\
&=& e^{q^\mu J_{\mu+}} e^{\eta^\alpha L_{\alpha+} - \xi^\alpha L_{\alpha+}} e^{\phi\tilde{J}_{+-}}G_0. \nonumber
\eea
Therefore we can see that
\[
\delta f = - \xi^\alpha \frac{\partial}{\partial \eta^\alpha} f + \ldots,
\]
and so we have the realisation
\beq
L_{\alpha+} = - \frac{\partial}{\partial \eta^\alpha}.
\eeq
Once again, scaling variables gives us $L_{\alpha+}$ in the momentum representation 
\[
L_{\alpha+} = -\lambda^{-1} \frac{\partial}{\partial \theta^\alpha}.
\]
Lastly, we shall calculate $L_{\mu\alpha}$; let $\rho = e^{\rho^{\mu\alpha} L_{\mu\alpha}}$, and so we have
\bea
\rho^{-1} x &=& e^{-\rho^{\mu\alpha} L_{\mu\alpha}} e^{q^\mu J_{\mu+} + \eta^\alpha L_{\alpha+}} e^{\phi\tilde{J}_{+-}}G_0, \nonumber \\
&=& e^{q^\mu J_{\mu+} - [ \rho^{\nu \alpha} L_{\nu\alpha}, q^\mu J_{\mu+} ]} e^{\eta^\alpha L_{\alpha+} - [ \rho^{\nu \alpha} L_{\nu\alpha}, \eta^\beta L_{\beta+}]} e^{\rho^{\mu \alpha} L_{\mu\alpha}} e^{\phi\tilde{J}_{+-}}G_0. \nonumber
\eea
To simplify this expression we use the commutation relations
\bea
q^\mu \rho^{\nu\alpha} [ J_{\mu+}, L_{\nu\alpha} ] &=& \rho^{\nu\alpha} q_\nu L_{\alpha+}, \nonumber \\
 -\eta^\beta \rho^{\nu\alpha} \{ L_{\nu\alpha},L_{\beta +} \} &=& \eta^\beta \rho^{\nu\alpha} \eps_{\alpha\beta} ( J_{\nu+} + a \eps^\mu_\nu J_{\mu +}), \nonumber \\
&=& - \rho^{\nu\alpha} \eta_{\alpha} J_{\nu +} - a \rho^{\mu\alpha} \eta_\alpha \eps_\mu^\nu J_{\mu+}. \nonumber 
\eea
thus we can write
\[
\rho f(q,\eta, \phi) = f( q^\mu - \rho^{\mu\alpha} \eta_\alpha - \rho^{\nu \alpha} a \eta_\alpha \eps^\mu_\nu, \eta^\alpha + \rho^{\nu\alpha} q_\nu, \phi),
\]
and so
\beq
L_{\nu\alpha} = - \eta_\alpha \left( \frac{\partial}{\partial q^\nu} + a \eps^\mu_\nu \frac{\partial }{\partial q^\mu} \right) + q_\nu \frac{\partial}{\partial \eta^\alpha}.
\label{d2eq:lma}
\eeq
Once again, changing to momentum representation and re-arranging gives us
\[
L_{\mu \alpha} = p_{\mu} \frac{\partial}{\partial \theta^{\alpha}}
- \theta_{\alpha} \frac{\partial}{\partial p^{\mu}} - a \theta_{\alpha}
\varepsilon_{\mu}{}^{\nu} \frac{\partial}{\partial p^{\nu}}. \\
\]

\section{Closure of algebra generated by $J_{MN}$}
\label{appenB}

Although the commutation relations of the generators $J_{MN}$ given in \eqn{eq:fulljs} must be equal to those given in \eqn{d2eq:secondjs} and \eqn{eq:homopart}, we shall test this by calculating $\{L_{\mu\alpha}, L_{\nu\beta} \}$ (We could choose any of the relations).

Equation \eqn{d2eq:secondjs} tells us that $ \{L_{\mu\alpha}, L_{\nu\beta} \} = \eps_{\alpha\beta} \eps_{\mu\nu} \tilde{J} - \eta_{\mu\nu} K_{\alpha\beta}$. Using the definition of $L_{\mu\alpha}$ given in \eqn{d2eq:lma} gives
\[
\begin{array}{l}
\{ - \eta_\alpha \partial_\mu - a \eta_\alpha \eps_\mu^\nu \partial_\sigma + q_\mu \partial_\alpha, - \eta_\beta \partial_\nu - a \eta_\beta \eps_\nu^\rho \partial_\rho + q_\nu \partial_\beta \} \\
= - \{ \eta_\alpha \partial_\mu, q_\nu \partial_\beta \} - a \{ \eta_\alpha \eps_\mu^\sigma \partial_\sigma, q_\nu \partial_\beta \} - \{ q_\mu \partial_\alpha, \eta_\beta \partial_\nu \} - a \{ q_\mu \partial_\alpha, \eta_\beta \eps^\rho_\nu \partial_\rho \}.
\end{array}
\]
Note that the first and third terms are identical, except for the indices, as are the second and fourth terms. Using the identity
\[
\{ AB,CD \} = \half \{ A,C \} \{ B,D \} + \half [A,C ][ B,D ],
\]
where $[A,B] = [C,D] = 0$, we have that the first term is
\[
-\half \eta_{\mu\nu} ( 2 \eta_\alpha \partial_\beta - \eps_{\alpha\beta} ) - \half \eps_{\alpha\beta} ( 2 q_\nu \partial_\mu + \eta_{\mu\nu} ).
\]
And so terms one and three sum to
\[
\eps_{\alpha\beta} ( q_\mu \partial_\nu - q_\nu \partial_\mu ) - \eta_{\mu\nu} ( \eta_\alpha \partial_\beta + \eta_\beta \partial_\alpha).
\]
In a similar fashion, we get that the second and fourth terms sum to 
\[
-a \eps_{\mu\nu} \eps_{\alpha\beta} \eta^\gamma \partial_\gamma - a \eps_{\alpha\beta} \eps_{\mu\nu} q^\rho \partial_\rho.
\]
Combining these two expressions together we get 
\bea
\{L_{\mu\alpha},L_{\nu\beta} \} &=& \eps_{\mu\nu} \eps_{\alpha\beta} \left( -q^\rho \eps_\rho^\sigma \partial_\sigma - a q^\rho \partial_\rho - a \eta^\gamma \partial_\gamma \right) - \eta_{\mu\nu} \left( \eta_\alpha \partial_\beta + \eta_\beta \partial_\alpha \right), \nonumber \\
&=& \eps_{\mu\nu} \eps_{\alpha\beta} \tilde{J} - \eta_{\mu\nu} K_{\alpha\beta},
\eea
as claimed.

\end{document}